\documentclass[5p]{elsarticle}

\usepackage{lineno}
\usepackage[breaklinks=true,unicode=true,urlcolor = blue,colorlinks = true,citecolor = blue,linkcolor = blue]{hyperref}
\usepackage{bm,latexsym,mathrsfs,enumerate,color}
\usepackage{amsmath,amssymb}
\renewcommand{\vec}[1]{\bm{#1}}

\journal{Journal of Magnetism and Magnetic Materials}

\begin{document}

\begin{frontmatter}

\title{Influence of Dzialoshinskii-Moriya interaction on static and dynamic properties of a transverse domain wall}

\author{Volodymyr P. Kravchuk\corref{mycorrespondingauthor}}
\address{Bogolyubov Institute for Theoretical Physics, 14-b, Metrologichna str., 03680 Kiev, Ukraine}
\cortext[mycorrespondingauthor]{Corresponding author}
\ead{vkravchuk@bitp.kiev.ua}

%
%

\begin{abstract}
It is shown that the Dzialoshinskii-Moriya interaction leads to asymmetrical deformation of the transverse domain wall profile in one-dimensional biaxial magnet. Amplitude of the deformation is linear with respect to the Dzialoshinskii constant $D$. Corrections caused by the Dzialoshinskii-Moriya interaction are obtained for the number of the domain wall parameters: energy density, D\"oring mass, Walker field. The modified $q$-$\Phi$ model with an additional pair of conjugated collective variables is proposed for studying the dynamical properties of the wall with taking into account the internal degrees of freedom.
\end{abstract}


\end{frontmatter}


\section{Introduction}
The influence of Dzialoshinskii-Moriya interaction~\cite{Dzyaloshinsky57} (DMI) on properties of domain walls (DW) is of high interest during the few last years. A especially drastic effect of DMI was recently reported for \textit{perpendicularly magnetized} ultrathin films: a large tilting of the DW surface was predicted under magnetic field or a spin polarized current driving~\cite{Boulle13a}, asymmetric expansion of the circular DW under field driving was demonstrated~\cite{Je13}. Moreover, new types of DW can appear, e.g. Dzialoshinskii DW~\cite{Thiaville12,Rohart13} and chiral DW~\cite{Emori13,Ryu13,Brataas13}. The last one demonstrates very high mobility under spin-transfer torques induced by spin Hall effect, this feature undoubtedly can be used for improvement of current controlled magnetic memory devices~\cite{Parkin08}.

This paper has two objectives: (i) to study the DMI influence on static and dynamical properties of a transverse DW in \textit{in-plane magnetized} narrow stripe, (ii) to propose the modified $q$-$\Phi$ model with an additional pair of canonically conjugated collective variables which enables one to study dynamics of the DW with taking into account the internal degrees of freedom.

First the collective variable approach was used to describe the dynamics of one-dimensional DW more then 40 tears ago~\cite{Slonczewski72,Thiele73}. Up to now the collective variable model proposed by Slonczewski~\cite{Slonczewski72,Malozemoff79} is widely used for different types of DWs and different drivings~\cite{Thiaville02a,Thiaville04,Thiaville05,Hillebrands06,Mougin07,Khvalkovskiy09,Landeros10,Thiaville12,Otalora12a,Otalora13} and it is often called ``$q$-$\Phi$ model''. In the frames of this model the DW dynamics is described by a pair of conjugated collective variables: $q$ determines position of the DW and angle $\Phi$ determines the magnetization orientation in the DW center. This model allows one to describe general properties of motion of the DW as a localized object, but it does not take into account internal degrees of freedom of the DW. However one should note a number of papers where the DW width $\Delta$ was considered as a third collective variable~\cite{Thiaville02a,Thiaville04,Landeros10,Hillebrands06}. But as it was shown in these works the width $\Delta$ is a slaved variable and its independent introduction to the model does not allow to describe the internal DW dynamics.

In this paper we show that due to the DMI the DW gains a deformation whose amplitude $\varkappa$ can be used as a collective variable conjugated to $\Delta$. In such a modified model the variables ($\Delta,\,\varkappa$) are not slaved and reflect a rich dynamics of the DW internal degrees of freedom.

The paper is organized as follows. In Section~\ref{sec:model} we introduce the model of long and thin stripe and discuss the considered interactions in the system. In Section~\ref{sec:statics} the static DW structure is obtained and deformations caused by DMI are considered. In the Section~\ref{sec:dynamics} the modified $q$-$\Phi$ model is introduced and DMI influence on the dynamical properties of the DW are studied. Results of the paper are summarized in the Section~\ref{sec:conclusions}.

\section{Model}\label{sec:model}
\begin{figure}
\center\includegraphics[width=0.8\columnwidth]{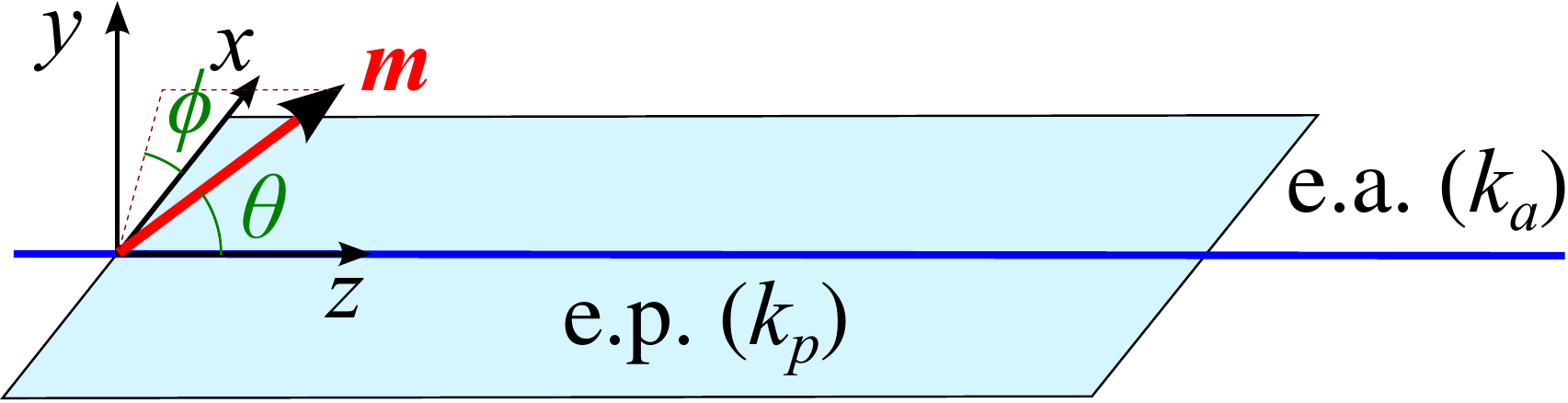}
\caption{Geometry and notations of the problem. The 1D biaxial ferromagnet with the easy-axis (e.a.) $z$ and easy plane (e.p) $z0x$ is considered. Angles $\theta$ and $\phi$ determine orientation of the magnetization vector $\vec m$.}\label{fig:geometry}
\end{figure}

We consider here a case of thin and narrow ferromagnetic stripe whose thickness and width are small enough to ensure the magnetization one-dimensionality, and the stripe length much exceeds the lateral dimensions. Thus the magnetization is described by continuous and normalized function $\vec m=\vec m(t,z)$, where $z$-axis is orientated along the stripe and $t$ denotes time. Since $|\vec m|=1$ it is convenient to proceed to the angular representation
 \begin{equation}\label{eq:theta-phi-parametrization}
  \vec m=(m_x,\,m_y,\,m_z)=(\sin\theta\cos\phi,\,\sin\theta\sin\phi,\,\cos\theta),
\end{equation}
where angle $\theta$ describes the deviation of magnetization from the magnet axis and angle $\phi$ determines orientation of vector $\vec m$ in plane perpendicular to $z$-axis. These notations are also explained in the Fig.~\ref{fig:geometry}. To describe the magnetization dynamics we use the phenomenological Landau-Lifshitz-Gilbert equations which in terms of angular variables have the form
\begin{subequations}\label{eq:LLG-angles}
  \begin{align}
  \label{eq:LLG-theta}-\sin\theta\dot{\theta}=&\frac{\delta\mathcal{E}}{\delta\phi}+\eta\sin^2\theta\dot\phi,\\
  \label{eq:LLG-phi}\sin\theta\dot{\phi}=&\frac{\delta\mathcal{E}}{\delta\theta}+\eta\dot\theta.
  \end{align}
\end{subequations}
Here the overdot indicates the derivative with respect to the dimensionless time $\tau=t\omega_0$, where $\omega_0=4\pi \gamma M_s$ with $\gamma$ being the gyromagnetic ratio and $M_s$ being the saturation magnetization and $\mathcal{E}=E/(4\pi M_s^2\mathcal{S})$ is the normalized energy with $\mathcal{S}$ being the area of the stripe cross-section. $\eta$ indicates the Gilbert damping. Typical scale of the physical quantities can be illustrated by the example of Permalloy: $\omega_0\approx30.3$~GHz, $4\pi M_s^2\approx0.93$~MJ/m$^3$, and $\eta\approx0.005-0.01$.

We assume that energy of the system has the following form
\begin{subequations}\label{eq:biax-E-field}
\begin{align}
  \label{eq:all-en-dens}  \mathcal{E}=\int\limits_{-\infty}^{\infty}[\mathscr{E}_{ex}+\mathscr{E}_{an}+\mathscr{E}_{DM}+\mathscr{E}_{z}]\mathrm{d}z,
  \end{align}
   The first term in \eqref{eq:all-en-dens} denotes the exchange energy
  \begin{align}\label{eq:Exchange}
  \mathscr{E}_{ex}=\frac{\ell^2}{2}(\partial_z\vec m)^2=\frac{\ell^2}{2}[\theta'^2+\sin^2\theta\phi'^2]
  \end{align}
   with $\ell$ being the exchange length and prime denotes  derivative with respect to the spatial coordinate $z$, e.g. $\theta'\equiv\partial_z\theta$.  The second term describes the anisotropy
  \begin{align}\label{eq:Anisotropy}
  \mathscr{E}_\mathrm{an}=&-\frac{k_a}{2}m_z^2+\frac{k_p}{2}m_y^2=\\ \nonumber
  =&\frac{k_a}{2}\sin^2\theta+\frac{k_p}{2}\sin^2\theta\sin^2\phi+\mathrm{const},
  \end{align}
  where $k_a>0$ and $k_p>0$ are constants of the easy-axis and easy-plane anisotropies respectively. The easy-axis is orientated along the stripe ($z$-axis) whereas the easy-plane coincides with the stripe plane ($x0z$), see Fig.~\ref{fig:geometry}. The anisotropy is chosen in such a form because for the thin and narrow stripe the expression~\eqref{eq:Anisotropy} approximately models the stray-field contribution originated from the surface magnetostatic charges\cite{Thiaville04,Mougin07}. Also the anisotropy term~\eqref{eq:Anisotropy} allows the head-to-head (tail-to-tail) transverse DW~\cite{Hillebrands06,Klaui08} which is the subject of this paper.

  Energy of the Dzialoshinskii-Moriya interaction (DMI) is taken in the form typical for cubic crystals with T symmetry~\cite{Bogdanov89,Cortes-Ortuno13}
  \begin{align}\label{eq:Dzialoshinskii}
  \mathscr{E}_{DM}=\frac{d}{2}\vec m\cdot[\nabla\times\vec m]=-\frac{d}{2}\sin^2\theta\phi',
  \end{align}
  where the normalized Dzialoshinskii constant $d=D/4\pi M_s^2$ is measured in units of length, therefore the comparison of $d$ and $\ell$ is equivalent to the comparison of strengths of the DMI and exchange contributions. The expression \eqref{eq:Dzialoshinskii} can be written in the equivalent form $\mathscr{E}_{DM}=\frac{\vec d}{2}[\vec m\times\partial_z\vec m]$ where the Dzialoshinskii vector $\vec d=-d\hat{\vec z}$ is collinear with the stripe; the case when the Dzialoshinskii vector is perpendicular to the one-dimensional magnet is considered in detail in Ref.~\cite{Heide11}.

  The last term
  \begin{align}
  \mathscr{E}_\mathrm{z}=-\vec h\cdot\vec m=-h\cos\theta
  \end{align}
  corresponds to interaction with the external magnetic field $\vec h=\hat{\vec{z}}H/4\pi M_s$, which is measured in units of the saturation field and is applied along the stripe.
\end{subequations}
\section{Static domain wall}\label{sec:statics}
In this section the no-driving case $h=0$ is considered and all the following analysis is based on the static form of equations~\eqref{eq:LLG-angles}:
\begin{subequations}\label{eq:stat-eqs}
\begin{align}
\label{eq:phi-stat}&\ell^2(\sin^2\theta\phi')'=k_p\sin^2\theta\sin\phi\cos\phi+d\sin\theta\cos\theta\theta'\\
\label{eq:theta-stat}&\ell^2\theta''=\sin\theta\cos\theta\left(\ell^2\phi'^2+k_a+k_p\sin^2\phi-d\phi'\right).
\end{align}
\end{subequations}
Without DMI ($d=0$) the ground state of the system with energy \eqref{eq:biax-E-field} is doubly degenerated: $\theta=0$ and $\theta=\pi$. The transition between domains of different ground states forms a DW. Structure of the DW can be found as a solution of \eqref{eq:stat-eqs} with boundary conditions $\theta(-\infty)=0$ and $\theta(+\infty)=\pi$ (case of head-to-head DW). This solution is well known~\cite{Landau35}
\begin{equation}\label{eq:DW-no-d}
\theta^{\mathrm{dw}}(z)=2\arctan\,e^{z/\Delta},\qquad\phi^{\mathrm{dw}}=0,
\end{equation}
where $\Delta=\ell/\sqrt{k_a}$ is width of the static DW\footnote{The case of tail-to-tail DW $\theta^{\mathrm{dw}}(z)=2\arctan\,e^{-z/\Delta}$ which originates from the opposite boundary conditions $\theta(-\infty)=\pi$ and $\theta(+\infty)=0$ , as well as case $\phi^{\mathrm{dw}}=\pi$ is absolutely analogous to the considered one.}. Nevertheless the DW solution \eqref{eq:DW-no-d} does not satisfy the equations \eqref{eq:stat-eqs} in case $d\ne0$. In the following we obtain deformation of the DW solution~\eqref{eq:DW-no-d} induced by the DMI, considering the DMI as a small perturbation $d/\ell\ll1$. With this purpose we introduce small deviations from the non-perturbed solution~\eqref{eq:DW-no-d}
\begin{equation}\label{eq:deviations}
\theta=\theta^{\mathrm{dw}}+\vartheta,\qquad \phi=\phi^{\mathrm{dw}}+\varphi.
\end{equation}
Substituting now \eqref{eq:deviations} into \eqref{eq:phi-stat} and linearizing the obtained equation with respect to the deviations one obtains the following equation for the deviation $\varphi$
\begin{equation}\label{eq:varphi}
\begin{split}
&\varphi''(\zeta)-2\tanh\zeta\,\varphi'(\zeta)-\alpha\varphi(\zeta)+\beta\tanh\zeta=0,\\
&\zeta=\frac{z}{\Delta},\quad \alpha=\frac{k_p}{k_a},\quad \beta=\frac{d}{\ell\sqrt{k_a}}.
\end{split}
\end{equation}
The corresponding equation for $\vartheta$ originated from \eqref{eq:theta-stat} is homogenous one $\vartheta''(\zeta)+(2\,\mathrm{sech}^2\zeta-1)\vartheta(\zeta)=0$, and besides the trivial solution\footnote{The trivial solution means that correction $\vartheta$ is of higher order of smallness than linear one with respect to the perturbation: $\vartheta=o(d/\ell)$.} $\vartheta=0$ it also has a solution $\vartheta=\mathrm{d}\theta^{\mathrm{dw}}/\mathrm{d}\zeta=(\cosh\zeta)^{-1}$ which corresponds to the translational shift of the DW~\eqref{eq:DW-no-d}.

\begin{figure}
\center\includegraphics[width=0.75\columnwidth]{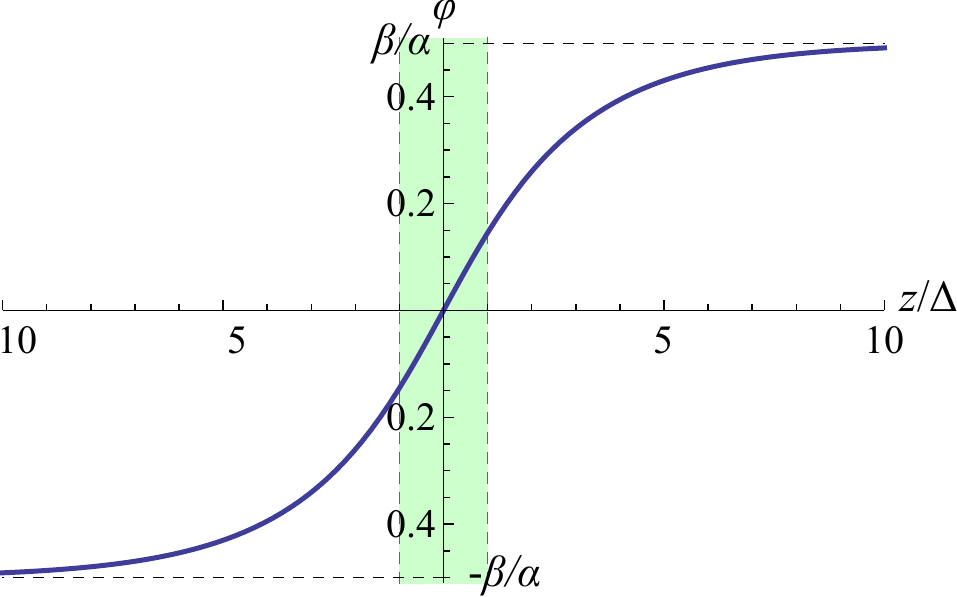}
\caption{The bounded solution of the Eq.~\eqref{eq:varphi} for parameters values $\alpha=1$, $\beta=0.5$. The shaded central region (green) shows size of the  DW width.}\label{fig:phi-tilde}
\end{figure}

Physically important solution of \eqref{eq:varphi} must be bounded one.  Such a solution for certain values of parameters is plotted in the Fig.~\ref{fig:phi-tilde}.
It exponentially approaches the horizontal asymptotes $\varphi=\pm\beta/\alpha$ and it is linear in the neighborhood of the origin
\begin{equation}\label{eq:varphi-0}
\varphi\approx\beta f(\sqrt{1+\alpha})\zeta,\qquad \zeta\rightarrow0,
\end{equation}
where function $f$ is determined in \eqref{eq:f}. See \ref{app:varphi} for details and the exact solution. It should be noted that though the correction $\varphi$ is not localized function the corresponding corrections for Cartesian components of the magnetization $m_i$ are localized within the DW width, for small $\varphi$
\begin{equation}\label{eq:mi}
  m_x\approx m_x^\mathrm{dw},\quad m_y\approx\frac{\varphi}{\cosh\zeta},\quad m_z\approx m_z^\mathrm{dw},
\end{equation}
where $m_i^\mathrm{dw}$ is the magnetization component of nonperturbed DW. Therefore the behavior of $\varphi$ at neighborhood of the origin \eqref{eq:varphi-0} only matters.

\begin{figure}
\includegraphics[width=\columnwidth]{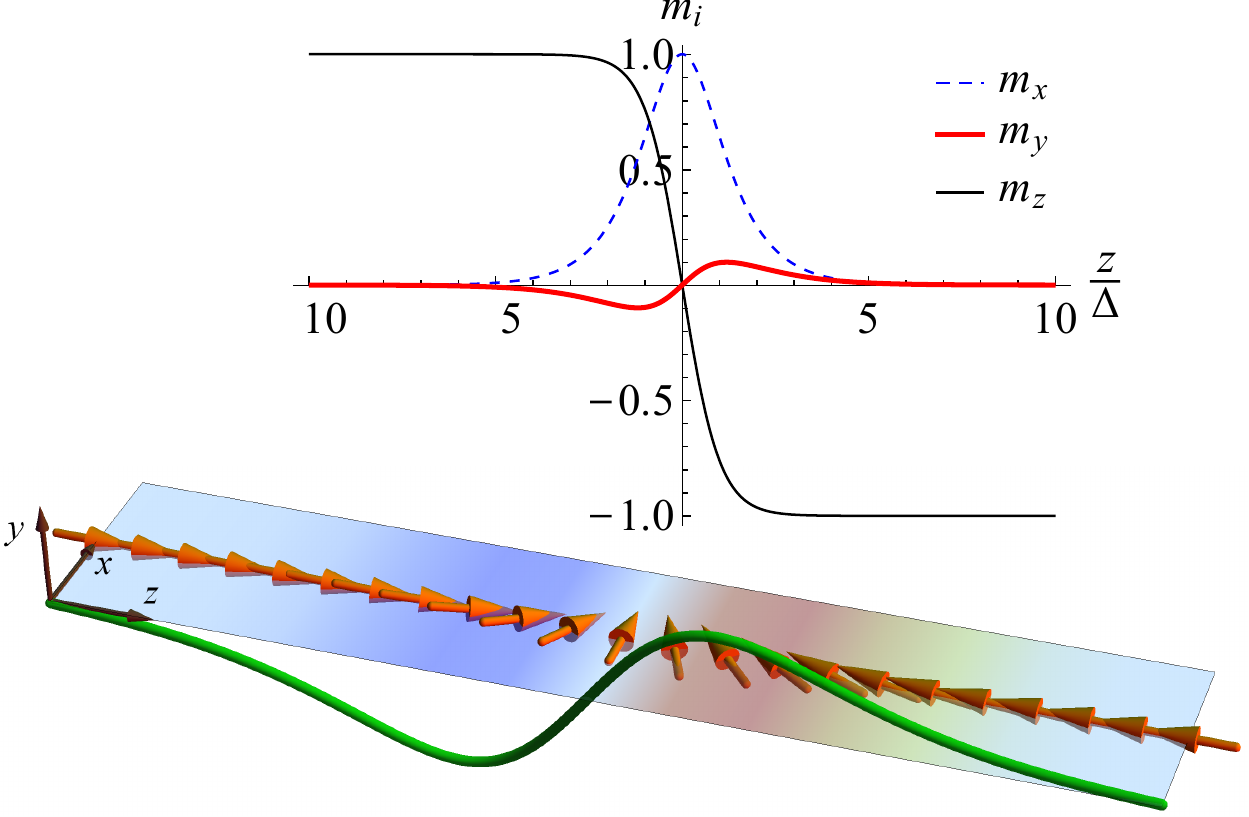}
\caption{Structure of the transverse DW deformed by the DMI. The top inset demonstrates the changing of the magnetization components along the stripe. Structure of the DM is shown below, distribution of the component $m_y$ is shown both by the stripe color and by 3D line. Parameters are the same as in the Fig.~\ref{fig:phi-tilde}}\label{fig:DW-structure}
\end{figure}

Structure of the DW, where the deformation $\varphi(\zeta)$ is taken into account, is shown in the Fig.~\ref{fig:DW-structure}. One should note the appearance of the out-of-plane component $m_y$ which is absent for the static transverse DW without DMI.

\section{Domain wall dynamics}\label{sec:dynamics}
To study dynamics of the DW we use the collective variable approach based on the Lagrangian formalism.  The equations of motion \eqref{eq:LLG-angles} can be treated as Euler-Lagrange equations
\begin{subequations}\label{eq:Euler-Lagrange}
  \begin{align}
  \label{eq:EL}\frac{\delta\mathcal{L}}{\delta\xi_i}-\frac{\mathrm{d}}{\mathrm{d}t}\frac{\delta\mathcal{L}}{\delta\dot{\xi}_i}&=\frac{\delta\mathcal{F}}{\delta\dot{\xi}_i},\qquad\xi=\theta,\,\phi
 \end{align}
 with the Lagrange function\footnote{Kinetic part of the Lagrange function is chosen in form convenient for the further integration with the Ansatz~\eqref{eq:ansatz}.}\cite{Doering48}
 \begin{align}
 \label{eq:Lagr-conven}\mathcal{L}=-\int\limits_{-\infty}^{\infty}\phi\sin\theta\dot\theta\mathrm{d}z-\mathcal{E},
 \end{align}
 and dissipative function
 \begin{align}\label{eq:diss-F}
  \mathcal{F}=\frac{\eta}{2}\int\limits_{-\infty}^{\infty}\left[(\dot\theta)^2+\sin^2\theta(\dot\phi)^2\right]\mathrm{d}z.
 \end{align}
\end{subequations}
Basing on the static solution \eqref{eq:DW-no-d} and taking into account the form of deformation \eqref{eq:varphi-0} we propose the following traveling wave Ansatz
\begin{equation}\label{eq:ansatz}
\begin{split}
&\theta(z,t)=2\arctan\,\exp\frac{z-q(t)}{\Delta(t)},\\
&\phi(z,t)=\Phi(t)+\varkappa(t)\frac{z-q(t)}{\Delta(t)},
\end{split}
\end{equation}
where ($q,\Phi$) and ($\Delta,\varkappa$) are two pairs of time dependent collective variables. Pair ($q,\Phi$) determines the general properties of motion of the DW as a localized object, whereas the pair ($\Delta,\varkappa$) describes the internal degrees of freedom. Substituting the Ansatz \eqref{eq:ansatz} into \eqref{eq:Lagr-conven} and performing the integration over $z$ one obtains the effective Lagrangian in form
\begin{equation}\label{eq:Lagr-eff}
\mathcal{L}=2(\Phi\dot q+c\varkappa\dot\Delta)-\mathcal{E},
\end{equation}
where $c=\pi^2/12$. Accordingly to \eqref{eq:Lagr-eff} the amplitude of the DW asymmetry $\varkappa$ is canonically conjugated momentum to the DW width $\Delta$, as well as DW phase $\Phi$ is canonically conjugated momentum to the DW position $q$. The effective energy in \eqref{eq:Lagr-eff} reads
\begin{equation}\label{eq:En-cv}
\begin{split}
\mathcal{E}&=\underbrace{\frac{\ell^2}{\Delta}(1+\varkappa^2)}_{\mbox{exchange}}+\underbrace{\Delta\left(k_a+k_p\sin^2\Phi+ck_p\varkappa^2\cos2\Phi\right)}_{\mbox{anisotropy}}\\
&\underbrace{-d\,\varkappa}_{\mbox{DMI}}\quad\underbrace{-2h\,q}_{\mbox{Zeeman}}.
\end{split}
\end{equation}
Here we assume that $\varkappa\ll1$ and therefore only terms linear and quadratic with respect to $\varkappa$ are considered.
The dissipative function can be obtained in the similar way by substituting the Ansatz \eqref{eq:ansatz} into \eqref{eq:diss-F}:
\begin{equation}\label{eq:F-diss-eff}
\mathcal{F}=\frac{\eta}{\Delta}\Bigl\{\dot q^2+(\dot q\varkappa-\dot\Phi\Delta)^2+c\left[\dot\Delta^2+(\dot\varkappa\Delta-\dot\Delta\varkappa)^2\right]\Bigr\}
\end{equation}
Lagrangian \eqref{eq:Lagr-eff} and dissipative function \eqref{eq:F-diss-eff} produce a set of equations of motion which in the lowest approximation with respect to $\varkappa$ can be written in form
\begin{subequations}\label{eq:cv-eqs-motion}
\begin{align}
\label{eq:q-Phi}\begin{Vmatrix}
\dot q/\Delta\\
\dot\Phi
\end{Vmatrix}=\,&\vec{\mathrm{M}}\begin{Vmatrix}
k_p\sin\Phi\cos\Phi\\
h
\end{Vmatrix},\\
\label{eq:kappa-Delta}2c\begin{Vmatrix}
\dot\Delta/\Delta\\
\dot\varkappa
\end{Vmatrix}=\,&\vec{\mathrm{M}}\begin{Vmatrix}
2\varkappa\left(\frac{\ell^2}{\Delta^2}+ck_p\cos2\Phi\right)-\frac{d}{\Delta}\\
\frac{\ell^2}{\Delta^2}-k_a-k_p\sin^2\Phi
\end{Vmatrix},
\end{align}
\end{subequations}
where matrix $\vec{\mathrm{M}}$ reads
\begin{equation}
\vec{\mathrm{M}}=\frac{1}{1+\eta^2}\begin{Vmatrix}
1-\eta\varkappa & \eta \\
-\eta & 1+\eta\varkappa
\end{Vmatrix}.
\end{equation}
Equations \eqref{eq:cv-eqs-motion} have the solution in form of translational motion $q=Vt$, $\Phi=\Phi_0$, $\Delta=\Delta_0$ and $\varkappa=\varkappa_0$. In a low damping case (one neglects terms $\eta^2$, $\eta\varkappa$ and $\eta d$) the parameters of this translational motion are the following
\begin{equation}\label{eq:dw-params}
\begin{split}
&\Phi_0=\frac12\arcsin\frac{h}{h_w^0},\quad h\le h_w^0=\eta k_p/2,\\
&\Delta_0=\ell(k_a+k_p\sin^2\Phi_0)^{-1/2},\\
&V= h\Delta_0/\eta,\\
&\varkappa_0=\frac{d\Delta_0}{2(\ell^2+ck_p\Delta_0^2\cos2\Phi_0)}.
\end{split}
\end{equation}

Thus in linear approximation with respect to $\varkappa$ the translational motion of the DW is the typical one for a biaxial magnet~\cite{Malozemoff79,Hillebrands06} except appearance of the asymmetrical deformation with amplitude $\varkappa_0$. Taking into account the neglected damping terms we were able to obtain the following correction for the Walker field
\begin{equation}\label{eq:hw-correction}
h_w\approx h_w^0\left(1-\frac{\eta d/\ell}{\sqrt{4k_a+2k_p}}\right).
\end{equation}
Since the obtained correction is linear with respect to $d$ and the Dzialoshinskii vector $\vec d$ is aligned along the magnet the Walker fields are expected to be slightly different for the opposite directions of the DW motion. The relative difference of the Walker fields is of order of magnitude $\sim\eta d$. It should be noted that the simplified form \eqref{eq:cv-eqs-motion} of the equations of motion is valid only under the condition $d/(\ell\sqrt{k_a})\ll1$, otherwise one should use an exact form of the equations of motion, see \ref{app:em-exact}.

Using \eqref{eq:dw-params} one can write the DW parameters $\Phi_0,\,\Delta_0,\,\varkappa_0$ as functions of the DW velocity. Then the substitution of this parameters into \eqref{eq:En-cv} enable us to series the DW energy in $V$:
\begin{subequations}
\begin{align}
\label{eq:E-ser-in-V}\mathcal{E}\approx\mathcal{E}_0\left(1-\frac{d^2}{\ell^2}\varepsilon_0\right)+\mathcal{M}\left(1-\frac{d^2}{\ell^2}\mu_0\right)\frac{V^2}{2},
\end{align}
where
\begin{align}
\mathcal{E}_0=2\sqrt{k_a}\ell
\end{align}
is energy of the static DM without DMI,
\begin{align}
\mathcal{M}=\frac{2\sqrt{k_a}}{k_p\ell}
\end{align}
is the D\"oring mass\cite{Doering48} of the DM without DMI. The DMI reduces values of these quantities and the corresponding corrections are proportional to $d^2$. In contrast, the corresponding correction of energy of a Bloch wall is linear with respect to the DMI~\cite{Bogdanov94,Heide08,Rohart13}. Values of the energy and mass corrections are determined by constants $\varepsilon_0=1/[8(k_a+ck_p)]$ and $\mu_0=[k_a(4c-1)+ck_p]/[8(k_a+ck_p)^2]$.
\end{subequations}

To study dynamics of the DW the regime $h>h_w$ (non-translational motion) we analyse the system \eqref{eq:cv-eqs-motion} numerically. First of all it should be noted that in the low damping case the Eqs.~\eqref{eq:q-Phi} coincide with the well known equations of the $q$-$\Phi$ model~\cite{Malozemoff79,Thiaville02a}. Therefore the general properties of the DW motion are close to ones obtained from the conventional $q$-$\Phi$ model: above the Walker breakdown the DW demonstrates oscillation motion with non-zero averaged velocity, at the same time the magnetization angle $\Phi$ precesses non-uniformly in time, for details see e.g.~\cite{Malozemoff79,Hillebrands06}.

A new feature of the proposed modification of the $q$-$\Phi$ model is internal DW dynamics which is described by the pair of conjugated variables $(\Delta,\varkappa)$, it is illustrated in the corresponding phase diagram, see Fig.~\ref{fig:circles}.
\begin{figure}
\includegraphics[width=\columnwidth]{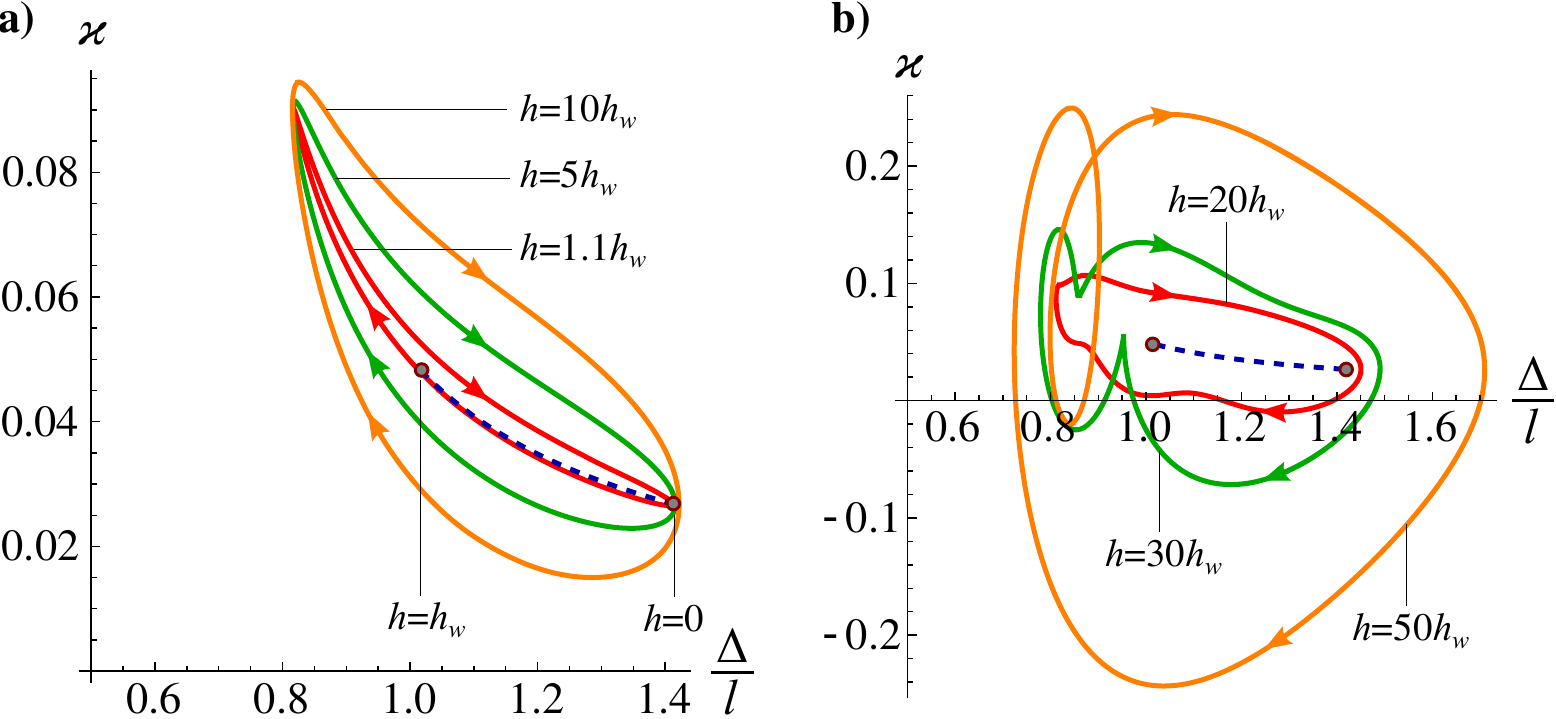}
\caption{Phase diagram for the conjugated pair $(\Delta,\varkappa)$. Dashed line is a locus of stationary points when the field is changing in the interval $0\le h<h_w$. The limit circles which appear for fields $h>h_w$ are shown by solid lines. Parameters are the following: $k_p=1$, $k_a=0.5$, $\eta=0.01$, $d=0.1\ell$.}
\end{figure}\label{fig:circles}
If the applied field is lower than the Walker breakdown $h<h_w$ then the single stationary point with coordinates $(\Delta_0,\,\varkappa_0)$ appears in the phase diagram. As the field increases from the value $h=0$ to the value $h=h_w$ the stationary point moves along the trajectory which is shown by dashed line in the Fig.~\ref{fig:circles}. Above the Walker breakdown $h>h_w$ the limit circle appears instead of the stationary point. The narrow limit circle appears abruptly with the finite size when the field overcomes the critical value $h=h_w$. The further field increasing leads to the limit circle broadening, see Fig.~\ref{fig:circles}a. When the applied field exceeds some critical value (for parameters used in the Fig.~\ref{fig:circles} it is $h_c\approx17h_w$) the amplitude $\varkappa$ of the DMI deformation changes its sign during the dynamics. At the same time the limit circle gains significant deformations with necks and loops, see Fig.~\ref{fig:circles}. It indicates the possibility of chaotic dynamics for the large applied fields. However, this issue is beyond the scope of the current paper and it should be studied in a separate work.

\section{Conclusions}\label{sec:conclusions}
It is shown that the DMI results in asymmetrical deformation of the profile of transverse DW. Amplitude of the deformation is linear with respect to the Dzialoshinskii constant $d$. To study dynamical properties of the DW the $q$-$\Phi$ model was modified by adding new pair of the conjugated collective variables (DW width, amplitude of the DMI deformation). That enables one to study the dynamics of internal degrees of freedom of DW and to find out caused by DMI corrections for the dynamical properties: Walker field gains the relative correction linear with respect to $d$, see \eqref{eq:hw-correction}; D\"{o}ring mass and energy of the static DW gain the corrections quadratic with respect to $d$, see \eqref{eq:E-ser-in-V}. The possibility of chaotic dynamics for high applied fields is indicated.

\section*{Acknowledgements}
The author is grateful to Prof. Yuri Gaididei and Prof. Franz Mertens for fruitful discussions.

\appendix
\section{Solution for the correction $\varphi$}\label{app:varphi}
Let us start from definition of the boundary conditions for the Eq.~\eqref{eq:varphi}. First of all, note that the Eq.~\eqref{eq:varphi} has an odd solution: $\varphi(-\zeta)=-\varphi(\zeta)$. Along with this, the bounded solution of the Eq.~\eqref{eq:varphi} has the following asymptotic behavior
\begin{equation}\label{eq:varphi-inf}
\varphi\approx \beta/\alpha-\mathcal{C}\exp[-(\sqrt{\alpha+1}-1)\zeta],\qquad\zeta\to\infty,
\end{equation}
where $\mathcal{C}>0$. This asymptotics  can be easily obtained as a solution of the simple linear ODE which appears from \eqref{eq:varphi} after replacement $\tanh\zeta\to1$. Basing on these two statements we can restrict ourselves with the solution of the Eq.~\eqref{eq:varphi} for interval $0\le\zeta<\infty$ with the following boundary conditions
\begin{equation}\label{eq:varphi-bc}
\varphi(0)=0,\qquad\varphi(\infty)=\beta/\alpha.
\end{equation}
Transition to the new function $u=\varphi/\cosh\zeta$ allow us to exclude the first derivative in the Eq.~\eqref{eq:varphi} and finally we proceed to the problem
\begin{subequations}\label{eq:u}
\begin{align}
  \label{eq:eq-u}&\frac{\mathrm{d}^2u}{\mathrm{d}\zeta^2}+\left(\frac{2}{\cosh^2\zeta}-\mu^2\right)u+\beta\frac{\sinh\zeta}{\cosh^2\zeta}=0,\\
  \label{eq:bc-u}&u(0)=0,\qquad u(\infty)=0,
\end{align}
\end{subequations}
where $\mu=\sqrt{1+\alpha}$. Since the change of variable $\xi=\tanh\zeta$ transforms the Eq.~\eqref{eq:eq-u} to an inhomogeneous Legendre equation
\begin{equation*}\label{eq:Legendre}
  (1-\xi^2)\frac{\mathrm{d}^2u}{\mathrm{d}\xi^2}-2\xi\frac{\mathrm{d}u}{\mathrm{d}\xi}+\left(2-\frac{\mu^2}{1-\xi^2}\right)u=-\beta\frac{\xi}{\sqrt{1-\xi^2}}
\end{equation*}
the general solution of \eqref{eq:eq-u} can be written as
\begin{equation}\label{eq:u-general}
\begin{split}
&u(\zeta)=C_1P_1^\mu(\tanh\zeta)+C_2Q_1^\mu(\tanh\zeta)+\beta\frac{\mu-1}{\mu+1}\frac{\Gamma(1-\mu)}{\mu\Gamma(\mu)}\times\\
&\times\!\!\!\int\limits_0^\zeta\!\!\frac{\left[Q_1^\mu(\tanh\zeta)P_1^\mu(\tanh t)-P_1^\mu(\tanh\zeta)Q_1^\mu(\tanh t)\right]\sinh t\,\mathrm{d}t}{\cosh^2t},
\end{split}
\end{equation}
where $P_1^\mu(x)$ and $Q_1^\mu(x)$ are Legendre functions of the first and second kinds respectively~\cite{NIST10}, $\Gamma(\mu)$ denotes the Gamma function and $C_1$ and $C_2$ are constants of integration. Applying now the boundary conditions \eqref{eq:bc-u} and using the corresponding asymptotic properties of the Legendre functions~\cite{NIST10} we obtain the following values of the integration constants
\begin{subequations}\label{eq:constants}
\begin{align}
\label{eq:C1-C2}&C_1=\frac{\beta\pi\cot\frac{\mu\pi}{2}f(\mu)}{2(\mu+1)\Gamma(\mu)},\quad C_2=-\frac{\beta f(\mu)}{(\mu+1)\Gamma(\mu)},\\
\label{eq:f}&f(\mu)=\frac{1}{2}-\frac{\mu-\frac1\mu}{4}\left[\psi\left(\frac{1-\mu}{4}\right)-\psi\left(\frac{3-\mu}{4}\right)+\frac{2\pi}{\cos\frac{\mu\pi}{2}}\right],
\end{align}
\end{subequations}
where $\psi(x)=\Gamma'(x)/\Gamma(x)$ is digamma function.

So solution of the Eq.~\eqref{eq:varphi} reds $\varphi(\zeta)= u(\zeta)\cosh\zeta$, where function $u(\zeta)$ is defined in \eqref{eq:u-general} and \eqref{eq:constants}. It is easy tosee now that $\varphi'(0)=u'(0)=\beta f(\mu)$, and therefore the asymptotic behavior \eqref{eq:varphi-0} takes place.

\section{Exact form of the equations of motion}\label{app:em-exact}
Substituting the Ansatz \eqref{eq:ansatz} into \eqref{eq:biax-E-field} and performing the integration over $z$ one obtains the effective energy in form
\begin{equation}\label{eq:En-exact}
\begin{split}
\mathcal{E}&=\underbrace{\frac{\ell^2}{\Delta}(1+\varkappa^2)}_{\mbox{exchange}}+\underbrace{k_a\Delta+\frac{k_p\Delta}{2}\left[1-\frac{\pi\varkappa}{\sinh(\pi\varkappa)}\cos2\Phi\right]}_{\mbox{anisotropy}}-\\
&\underbrace{-d\,\varkappa}_{\mbox{DMI}}\quad\underbrace{-2h\,q}_{\mbox{Zeeman}},
\end{split}
\end{equation}
which in harmonic approximation with respect to $\varkappa$ coincides with \eqref{eq:En-cv}. Lagrangian \eqref{eq:Lagr-eff} and dissipative function \eqref{eq:F-diss-eff} produce the following set of equations of motion
\begin{subequations}\label{eq:cv-eqs-motion-exact}
\begin{align}
&\frac{\dot q}{\Delta}-k_p\frac{\pi\varkappa}{\sinh(\pi\varkappa)}\sin\Phi\cos\Phi=\eta\left(\dot\Phi-\varkappa\frac{\dot q}{\Delta}\right),\\
&\dot\Phi-h=\eta\left[\varkappa\dot\Phi-\frac{\dot q}{\Delta}(1+\varkappa^2)\right],\\
&c\frac{\dot\Delta}{\Delta}-\frac{\ell^2}{\Delta^2}\varkappa+\frac{d}{2\Delta}+\frac{\pi}{4}k_p\cos(2\Phi)\frac{\sinh(\pi\varkappa)-\pi\varkappa\cosh(\pi\varkappa)}{\sinh^2(\pi\varkappa)}\\ \nonumber &=\eta c\left(\dot\varkappa-\varkappa\frac{\dot\Delta}{\Delta}\right),\\
&2c\dot\varkappa-\frac{\ell^2}{\Delta^2}(1+\varkappa^2)+k_a+\frac{k_p}{2}\left[1-\frac{\pi\varkappa}{\sinh(\pi\varkappa)}\cos2\Phi\right]\\ \nonumber
&=2\eta c\left[\varkappa\dot\varkappa-\frac{\dot\Delta}{\Delta}(1+\varkappa^2)\right],
\end{align}
\end{subequations}
which in linear approximation with respect to $\varkappa$ coincides with \eqref{eq:cv-eqs-motion}.



\end{document}